\documentclass[prb,preprint]{revtex4-1} 
\usepackage{amsmath}  % needed for \tfrac, \bmatrix, etc.
\usepackage{amsfonts} % needed for bold Greek, Fraktur, and blackboard bold
\usepackage{graphicx} % needed for figures

\begin{document}

% Be sure to use the \title, \author, \affiliation, and \abstract macros
% to format your title page.  Don't use lower-level macros to  manually
% adjust the fonts and centering.

\title{Hydrostatic-Based Proofs in Geometry}
% In a long title you can use \\ to force a line break at a certain location.

%When submitting the manuscript for review, do not include the author's name or institution
%\author{Daniel V. Schroeder}
%\email{dschroeder@weber.edu} % optional
%\altaffiliation[permanent address: ]{101 Main Street, Anytown, USA} % optional second address
% If there were a second author at the same address, we would put another 
% \author{} statement here.  Don't combine multiple authors in a single
% \author statement.
%\affiliation{Department of Physics, Weber State University, Ogden, UT 84408-2508}
% Please provide a full mailing address here.

%\author{David P. Jackson}
%\email{ajp@dickinson.edu}
%\affiliation{Department of Physics, Dickinson College, Carlisle, PA 17013}

% See the REVTeX documentation for more examples of author and affiliation lists.

\date{\today}

\begin{abstract}
Inspired by Tokieda’s work on mechanical insights in geometry, 
we explore a hydrostatic approach to classical geometric problems. Using principles of fluid statics, we derive two mathematical results regarding polygons. These results illustrate how physical principles can assist in understanding mathematical identities.
\end{abstract}

% AJP requires an abstract for all regular article submissions.
% Abstracts are optional for submissions to the "Notes and Discussions" section.

\maketitle % title page is now complete

\section{Introduction} % Section titles are automatically converted to all-caps.
% Section numbering is automatic.

The work of Tokieda, Levi, 
Vallejo \& Bove, and Kim shows how 
physical laws can lead to alternative proofs of mathematical theorems. \cite{tokieda:mechanical,levi:mathematical-mechanic,vallejo2024eppi,kim2025jensen}
In a similar 
spirit, in this paper we investigate how hydrostatic equilibrium can offer a novel approach to 
geometric problems.

We consider a system where rigid plates joined along their edges 
enclose a fluid. By analyzing the equilibrium of the fluid, we show that 
the polygon maximizing the enclosed area for fixed side lengths must have its 
vertices on a common circle; this is done in Section II. In Section III we derive the Shoelace formula 
for the area of a polygon by considering the forces acting on a submerged polygonal region. 
These results point out new connections between mathematics and physics.

\section{Isoperimetric Polygon Theorem}

We use the term ‘water’ to refer to an incompressible fluid of uniform density $\rho$. 
For simplicity, we set the atmospheric pressure to zero, although this assumption is 
not essential.

Among all simple closed polygons with given side lengths \( L_0, L_1, \cdots, L_{n-1} \) 
in counterclockwise order, the one enclosing the maximum area is cyclic; that is, all its 
vertices lie on a common circle. There are several mathematical proofs of this theorem, 
including approaches based on the isoperimetric inequality \cite{niven.lance:maxima} 
(Section 12.5), Bretschneider's formula \cite{crowdmath2018broken} (Lemma 2.4), 
and calculus methods \cite{peter2003maximizing, blasjo2005isoperimetric}. Here, we present a proof 
using hydrostatic principles.

\begin{figure}[h]
  \centering
  \includegraphics[trim=10pt 300pt -10pt 10pt, clip, width=0.7\textwidth]{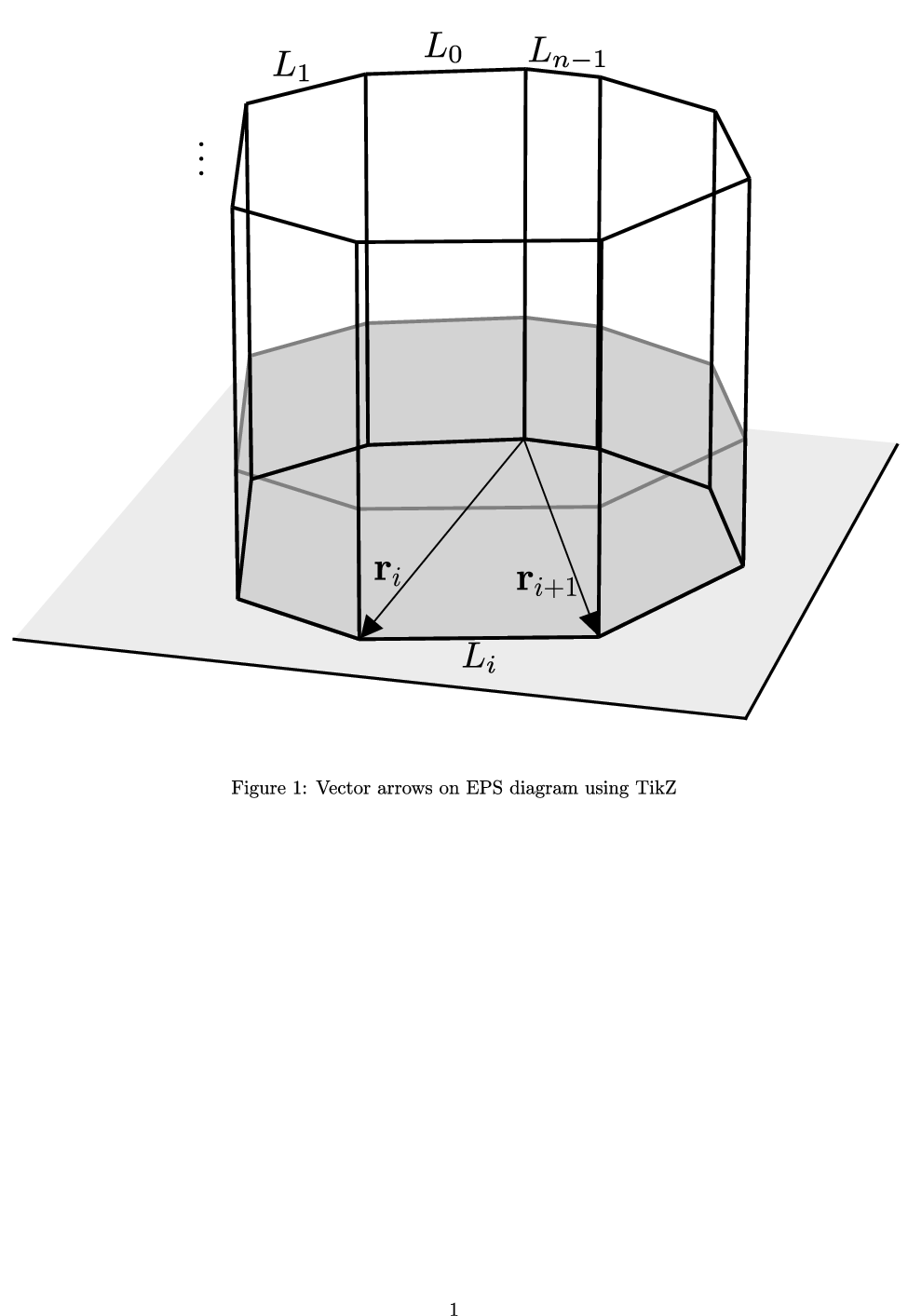}
  \caption{Picture of the fluid system}
  \label{fig:isoperpoly1}
\end{figure}

See Figure \ref{fig:isoperpoly1}. Consider $n$ rigid rectangular plates of widths $L_0, L_1, \cdots, L_{n-1}$. 
Join these plates along their edges to form a closed polygon, and then partially fill the enclosed region with water.
Assume that the plates can move freely. Since the volume of water is constant, its potential energy will be lowest when its area is highest, so the polygon's area will be maximized. (The existence of such a configuration is guaranteed by the compactness of the domain and the continuity of the area.)
Denote the position vectors of the polygon's vertices by $\mathbf{r}_0, \mathbf{r}_1, \cdots, \mathbf{r}_{n-1}$ in counterclockwise order, chosen so that the vector 
\begin{equation}\label{Li}
    \mathbf{L}_i=\mathbf{r}_{i+1}-\mathbf{r}_i
\end{equation}
has length $L_i$ for each $i$, as shown in Fig. 1. (We adopt the conventions $\mathbf{r}_i = \mathbf{r}_{i \bmod n}$.) We will refer to the plate corresponding to $\mathbf{L}_i$ as plate $P_i$.

Each plate exerts a force on its two neighboring plates. Figure 2 introduces the notation: $\mathbf{N}_i$ is the horizontal component(parallel to the ground) of the force exerted by plate $P_i$ on its adjacent plate $P_{i-1}$, for all $i$. By Newton’s third law, plate $P_{i-1}$ exerts an equal and opposite force on plate $P_i$; hence, the horizontal component of this reaction force is $-\mathbf{N}_i$, for all $i$.

At equilibrium, if the water is of depth $h$, the average  pressure on 
 plate $P_i$ is $p_{i,\text{av}}={\rho g h}/{2}$. 
Let $\mathbf{F}_i$ denote the force vector exerted by the water on plate $P_i$.
The magnitude of this force is 
\begin{equation}
    \|\mathbf{F}_i\| = p_{i,\text{av}} A_i = \frac{\rho g h^2}{2}L_i=fL_i,
\end{equation}
where we have defined $f = {\rho g h^2}/{2}$.  

\begin{figure}[h]
    \centering
    \includegraphics[trim=110pt 55pt 110pt 43pt, clip, width=0.5\textwidth]{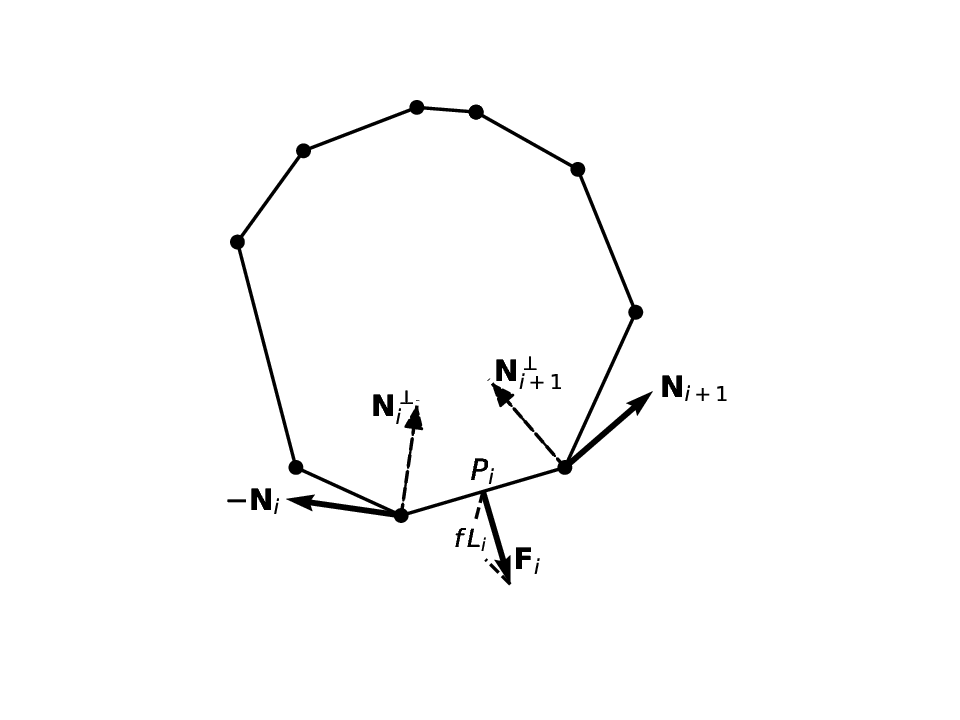}
    \caption{Force diagram of plate $P_i$}
    \label{fig:isoperpoly2}
\end{figure}

Since the system is in equilibrium, both the net force and the net torque on 
plate $P_i$ must vanish. In particular, the net force along the direction of $\mathbf{L}_i$ is zero, which implies that the components of the forces $\mathbf{N}_i$ and $\mathbf{N}_{i+1}$ in the direction of $\mathbf{L}_i$ are equal. Similarly, by taking the torque about the center of plate $P_i$, we deduce that the components of $-\mathbf{N}_i$ and $\mathbf{N}_{i+1}$ perpendicular to $\mathbf{L}_i$ must also be equal. Consequently, we obtain
\begin{equation}\label{eqforce}
    \|\mathbf{N}_i\| = \|\mathbf{N}_{i+1}\|
\end{equation}
for all $i$.

Define $\mathbf{N}_i^\perp$ as the vector obtained by rotating $\mathbf{N}_i$ by $90^\circ$ 
counterclockwise, for all $i$. When the force due to the water is rotated by $90^\circ$, it becomes $f\mathbf{L}_i$. Since the net force exerted on plate $P_i$ is zero, its 
rotated counterpart must also sum to zero. Consequently, 
\begin{equation}
    -\mathbf{N}_i^\perp +f \, \mathbf{L}_i + \mathbf{N}_{i+1}^\perp = 0.
\end{equation}
Dividing by $f$ and adding $\mathbf{r}_i+\mathbf{N}_i^\perp/f$ to both sides, and then using Eq. \eqref{Li}, this gives
\begin{align}
    \mathbf{r}_i+\frac{1}{f}\mathbf{N}_i^\perp &= \mathbf{r}_i+ \mathbf{L}_i+ \frac{1}{f}\mathbf{N}_{i+1}^\perp \\
    &= \mathbf{r}_{i+1} + \frac{1}{f}\mathbf{N}_{i+1}^\perp \label{centercommon}
\end{align}
for all $i$.

The equality in Eq. \eqref{centercommon} shows that $\mathbf{r}_i+\mathbf{N}_i^\perp/f$ is independent of $i$, and we denote this common vector as $\mathbf{r}_c$. It follows that
\begin{equation}
    \|\mathbf{r}_i-\mathbf{r}_c\| = \frac{1}{f}\|\mathbf{N}_i^\perp\| = \frac{1}{f}\|\mathbf{N}_i\|,
\end{equation}
and by Eq. \eqref{eqforce}, this length is identical for all $i$. Therefore, every vertex of the polygon lies on the circle centered at $\mathbf{r}_c$. This completes the proof.

\section{Shoelace Formula}

The Shoelace formula provides an easy method for computing the area of a simple polygon that is non-self-intersecting and encloses a single region without gaps given the Cartesian positions of its vertices.
Let $\mathcal{P}$ be an $n$-sided simple polygon that lies in the xy plane with vertices represented by position 
vectors $\mathbf{r}_0, \mathbf{r}_1, \cdots, \mathbf{r}_{n-1}$, arranged in counterclockwise order.   
If $\mathbf{r}_i = (x_i, y_i)$, then the area $A$ of the polygon is given by the Shoelace formula:
\begin{equation}
    A = \frac{1}{2} \sum_{i=0}^{n-1} (x_i y_{i+1} - x_{i+1} y_i).
\end{equation}

\begin{figure}[h]
    \centering
    \includegraphics[trim=80pt 120pt 80pt 140pt, clip, width=0.7\textwidth]{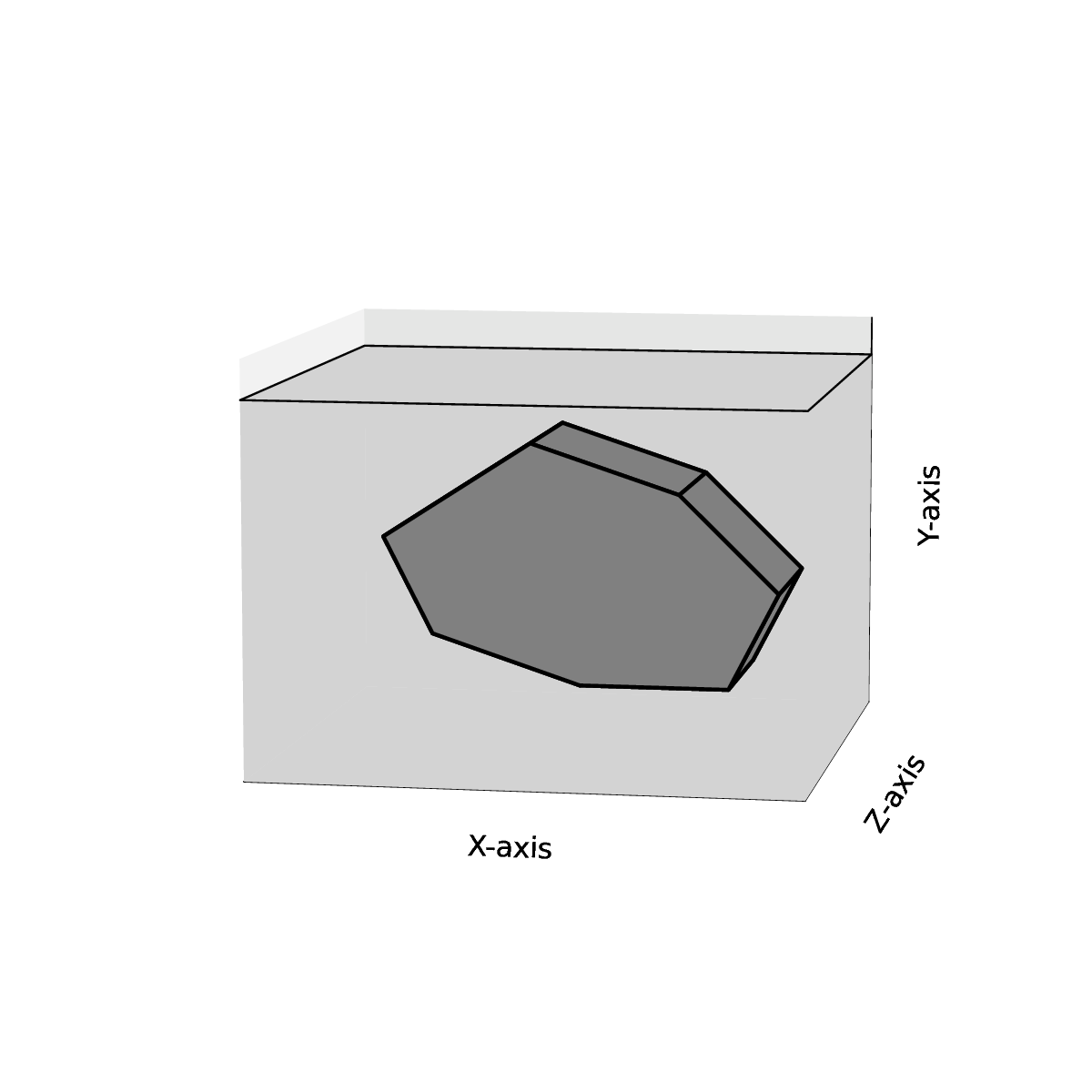}
    \caption{Prism region and water}
    \label{fig:shoelace1}
\end{figure}

We now develop a hydrostatic proof of this formula. Consider a solid prism that has the polygon as its base and has thickness $z_0$, as shown in Fig. \ref{fig:shoelace1}. It is submerged in water, with the top surface of the water in the plane $y=0$. (The submersion may require translating the prism in the negative $y$ direction.)
The gravitational acceleration is 
given by $\mathbf{g}=-g\mathbf{\hat{y}}$.

\begin{figure}[h]
    \centering
    \includegraphics[trim=60pt 30pt 60pt 50pt, clip, width=0.7\textwidth]{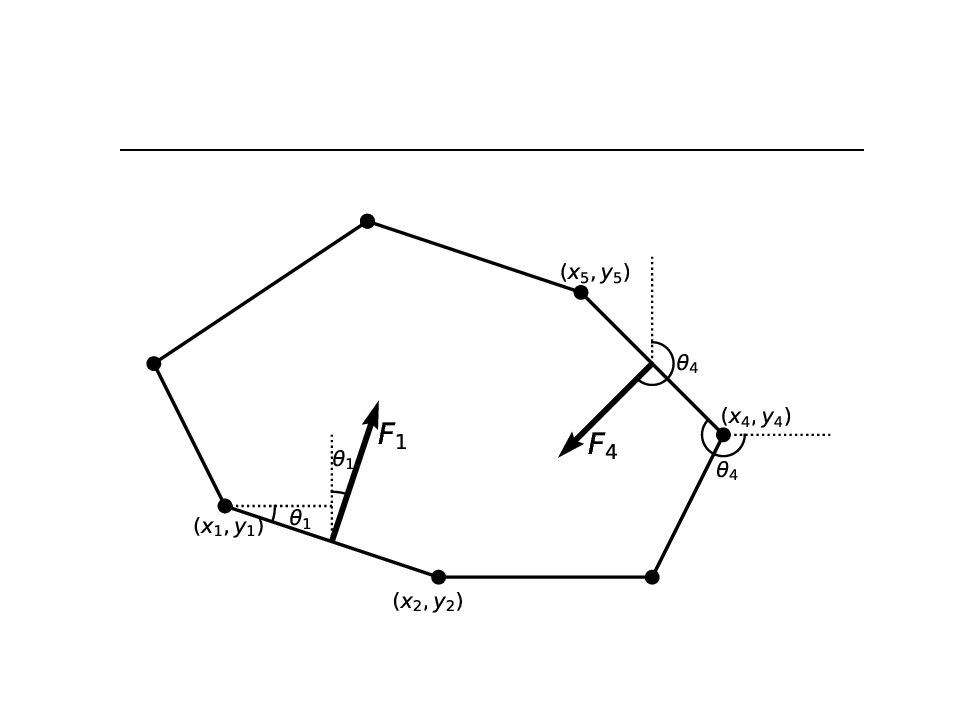}
    \caption{Force diagram for the hydrostatic pressure}
    \label{fig:shoelace2}
\end{figure}

If the system is in hydrostatic equilibrium, 
the net force acting on the water inside the prism must be zero.
The gravitational force acting on the water inside the prism is
\begin{equation}\label{grav}
    F_w = -\rho g z_0 A.
\end{equation}
Since the water pressure varies with depth, the average pressure on the face corresponding to the edge connecting the position vectors $\mathbf{r}_i$ and $\mathbf{r}_{i+1}$ is determined by the mean depth along that edge. This average pressure is
\begin{equation}
    p_{i, \text{av}} = \rho g \left(-\frac{y_i + y_{i+1}}{2}\right).
\end{equation}
Now see Fig. \ref{fig:shoelace2}. The force exerted on this face has magnitude
\begin{equation}
    F_i = p_{i, \text{av}}z_0 \|\mathbf{r}_{i+1}-\mathbf{r}_i\| = \rho g z_0 (-\frac{ y_i+y_{i+1}}{2}) \|\mathbf{r}_{i+1}-\mathbf{r}_i\|.
\end{equation}

The net vertical force resulting from the hydrostatic pressure, which corresponds to the buoyant force, is
\begin{align}
    F_b &= \sum_{i=0}^{n-1} F_i \cos\theta_i \nonumber \\
    &= \sum_{i=0}^{n-1} \rho g z_0(-\frac{ y_i+y_{i+1}}{2}) (x_{i+1} - x_i) \nonumber \\
    &= \frac{\rho g z_0}{2} \sum_{i=0}^{n-1} (x_i y_{i+1} - x_{i+1} y_i + x_i y_i - x_{i+1} y_{i+1}) \nonumber \\
    &= \frac{\rho g z_0}{2} \sum_{i=0}^{n-1} (x_i y_{i+1} - x_{i+1} y_i). \label{water}
\end{align}
The intermediate terms $x_iy_i-x_{i+1}y_{i+1}$ cancel out upon summation. Since the system is in equilibrium, 
\begin{equation}\label{tot}
    F_w+F_b = 0.
\end{equation}
This condition is, in fact, a direct expression of Archimedes' principle (see Ref. \cite{levi:mathematical-mechanic}, Sec.~2.7 for further discussion).
Combining Eqs. \eqref{grav}, \eqref{water}, and \eqref{tot}, we obtain
\begin{equation}
    A = \frac{1}{2} \sum_{i=0}^{n-1} (x_i y_{i+1} - x_{i+1} y_i).
\end{equation}
This completes the proof.

\section{Extensions to Continuous Systems}

\subsection{Dido's problem}
The classical optimization problem known as Dido’s problem asks which closed curve of fixed length $L$ encloses the maximum area. The isoperimetric polygon theorem provides a simple explanation: a closed curve can be viewed as a polygon with an infinite number of infinitesimal edges. For the enclosed area to be maximized, all vertices must lie on a common circle. Thus, the optimal shape must be a circle.

\subsection{Ideal gas law}
Now let \(\Omega\) represent a general three-dimensional submerged volume (not necessarily a prism). The gravitational force acting on the water within \(\Omega\) is
\begin{equation}
    \mathbf{F}_w = -\rho g \, \text{Vol}(\Omega)\, \hat{\mathbf{y}},
\end{equation}
and the buoyant force is given by
\begin{equation}
    \mathbf{F}_b = \iint_{\partial \Omega} \rho g (-y) (-d\mathbf{A}) = \rho g \iint_{\partial \Omega} y\, d\mathbf{A}.
\end{equation}

The condition that the net force on $\Omega$ vanishes in the hydrostatic state (i.e., Archimedes’ principle) leads to
\begin{equation}\label{archint}
    \rho g \iint_{\partial \Omega} y\, d\mathbf{A} = \rho g\, \text{Vol}(\Omega)\, \hat{\mathbf{y}}.
\end{equation}
This expression is often derived to justify Archimedes’ principle; see, for example, Ref.~\cite{lima2012archimedes}.
Canceling the common factor \(\rho g\) and taking the dot product with \(\hat{\mathbf{y}}\) yields
\begin{equation}
    \iint_{\partial \Omega} \mathbf{y} \cdot d\mathbf{A} = \text{Vol}(\Omega).
\end{equation}
By symmetry, it follows that
\begin{equation}
    \iint_{\partial \Omega} \mathbf{x} \cdot d\mathbf{A} =
    \iint_{\partial \Omega} \mathbf{y} \cdot d\mathbf{A} =
    \iint_{\partial \Omega} \mathbf{z} \cdot d\mathbf{A} =
    \text{Vol}(\Omega).
\end{equation}
Summing these equations gives
\begin{equation}\label{voldiv}
    \iint_{\partial \Omega} \mathbf{r} \cdot d\mathbf{A} =
    3\, \text{Vol}(\Omega),
\end{equation}
which is a special case of the divergence theorem.

Eq. \eqref{voldiv} is commonly used in the derivation of the ideal gas law \( PV = NkT \) for a gas confined within a container of arbitrary shape (see Example 7.14 in Ref.~\cite{marion2004classical}).  
Thus, it provides an alternative route to the ideal gas law that avoids explicit use of the divergence theorem, 
making the argument accessible to students unfamiliar with vector calculus.

\section{Discussion}
The approaches presented in this paper are not mathematically rigorous proofs in the traditional sense, but 
rather physically-motivated derivations. Interestingly, before considering the hydrostatic argument, the author was not even aware that the maximal-area polygon theorem existed. This demonstrates how physical reasoning can provide an intuitive bridge to mathematical concepts, making abstract results more accessible to students.

\section{Author Declarations}
The authors have no conflicts to disclose.


\begin{thebibliography}{99}

\bibitem{tokieda:mechanical} T. Tokieda, ``Mechanical Ideas in Geometry,''  
    \textit{Am. Math. Monthly} \textbf{105} (8), 697--703 (1998).  
    \url{https://www.tandfonline.com/doi/abs/10.1080/00029890.1998.12004951}.

\bibitem{levi:mathematical-mechanic} M. Levi,  
    \textit{The Mathematical Mechanic: Using Physical Reasoning to Solve Problems}  
    (Princeton University Press, Princeton, NJ, 2012).

\bibitem{vallejo2024eppi}
A. Vallejo and I. Bove,  
``Which is greater: $e^\pi$ or $\pi^e$? An unorthodox physical solution to a classic puzzle,"  
\textit{Am. J. Phys.} \textbf{92}, 397--398 (2024).  
\href{https://doi.org/10.1119/5.0188912}{https://doi.org/10.1119/5.0188912}.

\bibitem{kim2025jensen}
J. Kim,  
``Water-Based Proof of Jensen’s Inequality,"  
\textit{Am. Math. Monthly} \textbf{132} (2025), 1--5.  
\href{https://doi.org/10.1080/00029890.2025.2460413}{https://doi.org/10.1080/00029890.2025.2460413}.

\bibitem{niven.lance:maxima}I. Niven and L. H. Lance,  
    \textit{Maxima and Minima Without Calculus}  
    (American Mathematical Society, Providence, RI, 1981).

\bibitem{crowdmath2018broken}
P. A. CrowdMath,  
``The Broken Stick Project,"  
arXiv:1805.06512 (2018).  
\url{https://arxiv.org/abs/1805.06512}.

\bibitem{peter2003maximizing}
T. Peter,  
"Maximizing the Area of a Quadrilateral,"  
\textit{Coll. Math. J.} \textbf{34} (4), 315--316 (2003).  
\url{https://www.tandfonline.com/doi/abs/10.1080/07468342.2003.11922024}.

\bibitem{blasjo2005isoperimetric}
V. Blåsjö,  
``The evolution of the isoperimetric problem,''  
\textit{Am. Math. Monthly} \textbf{112}(6), 526--566 (2005).  
\href{https://doi.org/10.1080/00029890.2005.11920227}{https://doi.org/10.1080/00029890.2005.11920227}.

\bibitem{lima2012archimedes}
F. M. S. Lima,  
``Using surface integrals for checking Archimedes' law of buoyancy,"  
\textit{Eur. J. Phys.} \textbf{33}, 101--113 (2012).  
\href{https://doi.org/10.1088/0143-0807/33/1/009}{https://doi.org/10.1088/0143-0807/33/1/009}.

\bibitem{marion2004classical}
S. T. Thornton and J. B. Marion,  
\textit{Classical Dynamics of Particles and Systems}, 5th ed.  
(Brooks/Cole, Belmont, CA, 2004).

\end{thebibliography}
\end{document}